\documentclass{article}
\usepackage{amsfonts}
\usepackage{xcolor}
\usepackage{microtype}
\usepackage{graphicx}
\usepackage{subfigure}
\usepackage{booktabs} %
\usepackage{xspace}
\usepackage{multirow}
\usepackage{amsmath}
\usepackage{algorithmic}

\usepackage{threeparttable}
\usepackage{tabularx}
\usepackage{hyperref}
\usepackage{xurl}
\usepackage{comment}

\newcommand{\name}{Difflet\xspace}
\usepackage[accepted]{mlsys2025}
\usepackage{etoolbox}
\makeatletter
\renewcommand{\Notice@String}{}
\patchcmd{\printAffiliationsAndNotice}{\ \\}{}{}{\PackageWarning{difflet}{notice line not removed}}
\patchcmd{\printAffiliationsAndNotice}{Correspondence to: }{\newline}{}{\PackageWarning{difflet}{author footnote not patched}}
\patchcmd{\printAffiliationsAndNotice}{\hspace*{-\footnotesep}}{\raggedright\hspace*{-\footnotesep}}{}{\PackageWarning{difflet}{author footnote not ragged}}
\makeatother
\usepackage{comment}

\newlength{\figurecaptionskip}
\makeatletter
\def\difflet@figuretype{figure}
\long\def\@makecaption#1#2{%
  \ifx\@captype\difflet@figuretype
    \vskip\figurecaptionskip
  \else
    \vskip 0pt
  \fi
  \baselineskip 0pt
  \setbox\@tempboxa\hbox{#1. #2}%
  \ifdim\wd\@tempboxa>\hsize
    \sbox{\newcaptionbox}{\small\sl #1.~}%
    \newcaptionboxwid=\wd\newcaptionbox
    \usebox\newcaptionbox {\footnotesize #2}%
  \else
    \centerline{{\small\sl #1.} {\small #2}}%
  \fi
}
\makeatother

\definecolor{bin}{RGB}{200, 100, 50}

\mlsystitlerunning{\name: Diffusion Transformer Inference on AWS Trainium}

\AtBeginDocument{%
  \raggedbottom
  \setlength{\abovedisplayskip}{5pt}%
  \setlength{\belowdisplayskip}{5pt}%
  \setlength{\abovedisplayshortskip}{3pt}%
  \setlength{\belowdisplayshortskip}{3pt}%
}

\begin{document}
\twocolumn[

\mlsystitle{\name: A Fast Inference Engine for Diffusion Transformer Models on AWS Trainium}  %

\mlsyssetsymbol{equal}{*}

\begin{mlsysauthorlist}
\mlsysauthor{Bin Ma}{yotta}
\mlsysauthor{Min Yu}{yotta}
\mlsysauthor{Vijay Kodamalla}{yotta}
\mlsysauthor{Johnny Liu}{yotta}
\mlsysauthor{Da Li}{yotta}
\mlsysauthor{Dong Li}{yotta,ucm}
\end{mlsysauthorlist}

\mlsysaffiliation{ucm}{University of California, Merced, Merced, California, USA}
\mlsysaffiliation{yotta}{Yotta Labs, USA}

\mlsyscorrespondingauthor{Bin Ma}{bin@yottalabs.ai}
\mlsyscorrespondingauthor{Min Yu}{minyu@yottalabs.ai}
\mlsyscorrespondingauthor{Vijay Kodamalla}{vijayk@yottalabs.ai}
\mlsyscorrespondingauthor{Johnny Liu}{liu@yottalabs.ai}
\mlsyscorrespondingauthor{Da Li}{d@yottalabs.ai}
\mlsyscorrespondingauthor{Dong Li}{dli35@ucmerced.edu}

\mlsyskeywords{Machine Learning, MLSys}

\vskip 0.3in

\begin{abstract}
On AWS Trainium2, every component of a diffusion model runs as a program compiled ahead of time for fixed shapes, while pipelines differ across models, prompt lengths across requests, and caching decisions across denoising steps. We present \name, a serving engine in which five model pipelines share one compilation backend, joint attention is split across cores by attention head, and a host controller decides at each denoising step whether to run the denoiser or reuse its earlier output. A small probe program computes the caching signal on the device and returns one number per step, and a resident model process keeps every program loaded across requests. Across five image and video model families, fused attention accelerates denoiser steps by 1.2--4.3$\times$ over the same framework using PyTorch/XLA attention, and at 32.8k visual tokens it is the only attention that compiles. On four of the five models, resident serving is 1.5--3.4$\times$ faster than standalone runs that reload weights from a warm file cache for each generation.
\end{abstract}
]

\printAffiliationsAndNotice{}  %

\section{Introduction}
\label{sec:introduction}

Diffusion transformers generate images and videos through repeated denoising steps. Each step applies a transformer to a noisy latent representation, using the timestep and conditioning inputs to predict the noise, or for flow-matching models the velocity, from which the sampler computes a less noisy latent~\cite{dit}. Higher image resolutions and longer videos increase the number of visual tokens processed at each step. Repeating this computation raises both generation latency and the accelerator time consumed by each request. Efficient serving therefore requires shortening the user's wait while controlling the cost per generated image or video.

Most diffusion serving today runs on NVIDIA GPUs, whose high price and limited supply make the production cost hard to manage. Cloud providers have therefore built cost-efficient, customized accelerators. AWS Trainium~\cite{fan2024hlat} is one of those examples: it has been reported that Trainium2 instances provide 30--40\% better price-performance than GPU-based EC2 P5e and P5en instances~\cite{aws-trn2-instances}, and distributed LLM training on first-generation Trainium cost up to 46.3\% less than on A100-based instances in selected workloads, at comparable or higher training throughput~\cite{fu2024distributed}. Each Trainium chip contains several NeuronCores (cores for short), and the Neuron Kernel Interface (NKI) lets programmers write kernels that choose which of a core's tensor, vector, scalar, and general-purpose engines runs each operation and when data moves~\cite{aws-trn2-nki,aws-ncv3,aws-neuron}. With Trainium's production software stack, serving a diffusion model means running programs that are compiled ahead of time for fixed shapes and called by a host process at every denoising step. Serving diffusion models under this execution model requires deciding what must be fixed in compiled programs and what must remain flexible across models, requests, and denoising steps. This creates challenges at three levels: accommodating different model pipelines, distributing each denoiser evaluation across cores, and reusing results across denoising steps.

First, diffusion models differ in their encoders, denoiser, and decoder and in the order they call them, yet on Trainium every one of these components must be compiled before serving. Ahead-of-time (AOT) compilation specializes each component for the selected tensor shapes, data type, and parallel configuration~\cite{aws-neuronx-trace}. When compilation, loading, and invocation are embedded in each model's pipeline, supporting another model requires assembling another execution path around these constraints. The challenge is to add a model without writing another compilation and loading path for it.

Second, distributing a denoiser across Trainium cores requires finding a parallel configuration that both the model and the compiled kernels support. Tensor parallelism (TP) partitions weights~\cite{megatron}; sequence and context parallelism (SP, CP) partition tokens~\cite{megatronsp,ulysses,ringattention}; and classifier-free guidance (CFG) parallelism executes conditional and unconditional denoiser passes concurrently~\cite{cfg}. Each choice changes tensor shapes, data layouts, and communication requirements, so these choices cannot be combined freely. For example, joint attention over visual and text tokens must exclude padded text keys from its softmax. The masked attention path used here requires valid keys to be contiguous and query and key lengths to match~\cite{aws-nki-attention-cte}. A direct partition that keeps queries local while gathering all keys violates the second requirement. Such incompatibilities make parallel configuration a question of execution feasibility as well as performance, and discovering them only after compilation is costly.

Third, step caching, a common optimization for diffusion transformers, must obtain a cheap reuse signal despite Trainium's compilation and memory constraints. Methods such as TeaCache use early denoiser features to decide when to reuse prior results~\cite{teacache}. In our AOT execution path, these features remain inside the compiled denoiser, while the host controls whether to invoke it. Computing the signal separately on the host adds latency; a device implementation instead needs additional memory for weights and saved features alongside the partitioned denoiser. This placement tradeoff can undermine caching: our host path makes Wan's denoising loop 11\% longer than without caching, while HunyuanVideo already occupies 22 of 24\,GiB per logical core before adding a probe.

We present \name, a serving engine for diffusion transformers on AWS Trainium2 built around one boundary: what is the same for every request is compiled, and what changes per request or per step stays on the host. Each component's program fixes its shapes, data type, parallel configuration, and kernels; the model's control flow, the prompt length, the decision to skip a step, and each request's sampling state are left to the host. When such a value must reach or leave the device, \name passes it as a small runtime value instead of building a new program: the number of valid text keys going into attention, and one change score per step coming out of the caching probe.

To support different models, \name exploits what diffusion pipelines have in common. Their control flow differs, but all of them are built from conditioning encoders, a denoiser, and a decoder. Each of these components compiles to a program that depends only on its shape, data type, parallel configuration, and compiler version, and, where a device probe is fused into the denoiser, on whether the probe is present; runtime settings such as the caching cadence or threshold do not enter it. \name therefore keeps each model's control flow in its own pipeline on the host and moves everything AOT compilation needs into a shared backend, driven by a short description of each model. The backend caches each compiled program under exactly these inputs, so a server restart or a new caching threshold reuses the program instead of recompiling it.

For parallel execution, \name chooses each split by what Trainium's attention kernels accept. For attention, it uses Ulysses-style context parallelism~\cite{ulysses}, which regroups the tokens by attention head, so each core holds every query and key for its heads and thus the same number of queries as keys. \name places the text padding after all valid keys, so the valid keys sit next to each other, and each request passes only the number of valid keys. One compiled program thus serves every prompt length, and the split result matches unsplit attention to within BF16 rounding. \name also rejects unsupported combinations before compiling rather than after a long compilation fails.

To skip denoising steps, \name estimates the output change outside the denoiser in one of two ways. A small compiled probe computes an early denoiser feature on the device, keeps the previous step's feature there, and returns a single number to the host; because it loads only the weights of the layers it runs, it fits beside the denoiser even on HunyuanVideo. Alternatively, \name estimates the change from the outputs of earlier steps, which needs no additional program. In both cases, a host controller decides at each step whether to call the denoiser or reuse an earlier output.

These three parts meet in \name's resident model process, which keeps a model's programs loaded across requests. At startup, the process loads the programs compiled for the deployment's shapes, data type, and parallel configuration, checks that they match, and runs a test generation before accepting requests. For every request, it resets the sampling and caching state, so reuse never crosses requests. \name serves FLUX~\cite{fluxdev}, Qwen-Image~\cite{qwenimage}, Wan~\cite{wan}, HunyuanVideo~\cite{hunyuanvideo}, and LTX-2~\cite{ltx2} through this process.

We summarize our contributions as follows:
\begin{itemize}
    \item A shared Trainium backend through which one kernel or parallel mode serves different diffusion pipelines.
    \item Parallel execution planned as a serving decision, with split joint attention that matches unsplit attention.
    \item Step caching around compiled denoisers, with or without a device probe.
    \item A resident model process that serves five image and video models.
\end{itemize}

\section{Background}

\textbf{Trainium2 architecture.} AWS Trainium2 has eight physical NeuronCore-v3 cores and 96\,GiB of high-bandwidth memory (HBM) for weights and activations~\cite{aws-trn2-arch}. Each core provides a software-managed scratchpad (SBUF) for tensors and an accumulation buffer (PSUM) for matrix multiplication results. Direct memory access (DMA) engines transfer tensors between HBM and SBUF~\cite{aws-trn2-nki}. The Tensor Engine handles matrix multiplication, Vector and Scalar handle normalization and activations, and GpSimd supports programmable SIMD. These engines can execute concurrently and overlap computation with data movement~\cite{aws-ncv3,aws-trn2-nki}. Collective-communication cores support all-reduce and all-to-all~\cite{aws-trn2-nki}. We use the default $\mathrm{LNC}{=}2$ configuration, which pairs physical cores into four logical cores per device, each with 24\,GiB of HBM; all parallel configurations count logical cores~\cite{aws-lnc}.

\textbf{Compilation and execution.} The Neuron SDK provides a compiler, runtime, and Neuron Kernel Interface (NKI) for custom kernels with explicit data movement~\cite{aws-neuron}. Our ahead-of-time workflow compiles PyTorch components for fixed tensor shapes and data types; padding or compiled variants support different input sizes~\cite{aws-neuronx-trace}. The runtime loads programs onto NeuronCores, while host code invokes them and controls iteration and data-dependent decisions.

\section{Model Pipelines and a Shared Trainium Backend}
\label{sec:frontend}

\begin{figure}[t]
    \centering
    \includegraphics[width=\columnwidth]{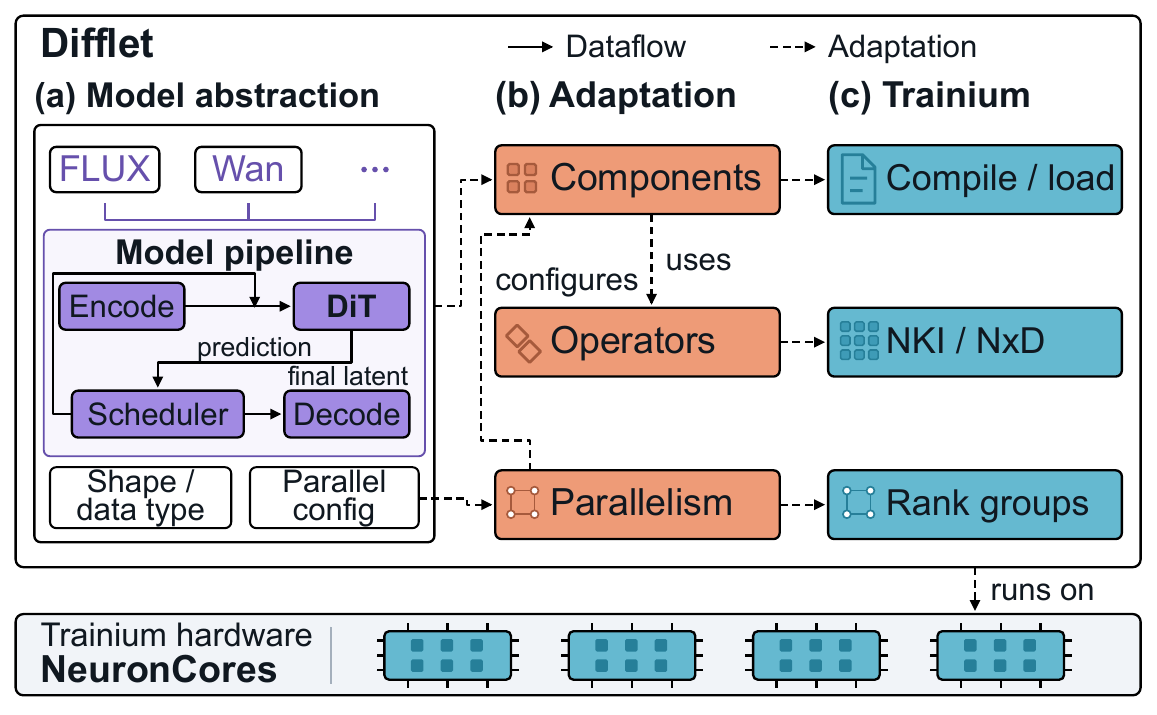}
    \caption{Model abstraction and Trainium adaptation in \name. (a) Pipelines specify component dataflow and configurations. (b) Component, operator, and parallelism interfaces connect models to (c) Trainium compilation and execution. Solid arrows denote dataflow; dashed arrows denote configuration and adaptation. NeuronCore icons are schematic.}
    \label{fig:model-abstraction-trainium}
\end{figure}

\name splits each model along the boundary of Section~\ref{sec:introduction}: a pipeline on the host decides which components run and in what order, and a shared Trainium backend compiles, loads, and calls them (Figure~\ref{fig:model-abstraction-trainium}). The pipeline is hardware-agnostic: it names the model's encoders, denoiser, and decoder, the tensors that flow between them, and the shapes each one takes. A short model adapter connects the two: given a resolution, data type, and parallel configuration, it tells the backend which components to compile and how to shard each one.

The model pipeline determines how tensors move through these components. For a generation request, conditioning encoders produce the context used during denoising, and the pipeline prepares the initial latent and timestep schedule. At each step, the denoiser predicts noise or a velocity field from the current latent, timestep, and encoder outputs. The diffusion scheduler uses this prediction to update the latent for the next step. After the final update, the decoder maps the latent to an image or video. The pipeline owns this iterative control flow, including model-specific conditioning and component choices.

Trainium requires components to be compiled for static tensor shapes. Before inference, \name therefore prepares the required components for the selected image resolution or video length, data type, and parallel configuration. The Trainium backend uses this information to compile each component ahead of time, verify that the required compiled artifacts are present and match the selected configuration, and load the programs and weights onto the assigned cores. These preparation steps are separate from the generation process: loading order can follow resource requirements while the model pipeline retains the model's execution order.

During inference, the model pipeline runs in Python and calls the components' programs with tensor inputs, receiving output tensors for the next stage or denoising step. Within each component, shared operator interfaces connect model operations to Trainium implementations, including NKI kernels and NeuronX Distributed (NxD) operations. The parallel configuration specifies how work is distributed across ranks and is checked against supported combinations. The backend organizes these ranks into groups for the selected parallel modes. Each model thus keeps its own control flow while sharing one backend. Because models call attention and collectives through shared operator interfaces, a kernel or parallel mode added once becomes available to every model that uses that interface.

\section{Parallel Execution on Trainium}
\label{sec:parallel-execution}

\name distributes each denoiser evaluation across cores with tensor, sequence, context, CFG, and data parallelism; each program fixes its parallel configuration, while the prompt length remains a runtime value. Each rank is one logical core, which pairs two physical NeuronCore-v3s under the default $\mathrm{LNC}{=}2$ configuration. On the single-chip \texttt{trn2.3xlarge}, collectives between ranks stay on chip, with DMA engines moving data between the logical cores' HBM~\cite{aws-lnc,aws-trn2-nki}. Within a rank, each attention kernel runs on both physical cores, which split the call's heads between them~\cite{aws-nki-attention-source}. Each physical core computes on tiles in its SBUF, whose 128 partitions set the tile height, and its Tensor Engine is a $128\times128$ systolic array~\cite{aws-trn2-nki}. Two properties shape the design below: compiled programs fix every tensor shape, and kernels process sequences in tiles of 128 tokens, one per SBUF partition. In joint attention, each query attends to both visual and text keys under a single softmax normalization. Distributed execution must preserve this normalization while excluding padding and counting replicated text keys only once. These requirements determine how \name redistributes tensors and combines results from Trainium's fused attention kernels.

\subsection{Composing Parallelism}
\label{sec:parallel-composition}

\textbf{Tensor parallelism (TP)} partitions linear-layer weights across ranks using the Megatron column and row decomposition~\cite{megatron}. Ranks in a TP group process the same tokens with different weight slices; an all-reduce sums the partial outputs and returns the result to each rank. \textbf{Sequence parallelism (SP)} partitions tokens within that group for normalization, modulation, and residual operations between linear layers~\cite{megatronsp}. An all-gather assembles the token partitions before the linear layers, and a reduce-scatter sums and partitions their outputs. The normalization, modulation, and residual operations run on the Vector and Scalar Engines, whose throughput is far below the Tensor Engine's~\cite{aws-ncv3}; SP divides this otherwise replicated work by the TP size while reusing the TP group.

\textbf{Context parallelism (CP)} partitions tokens across a separate rank group. Ranks at the same TP position hold the same weight partition but process different tokens. HunyuanVideo and Qwen-Image partition visual tokens while replicating text; FLUX partitions both streams. We focus below on the replicated-text layout, also used in xDiT~\cite{xdit-text-replica}.

Each rank's queries (Q) require keys (K) and values (V) from all visual partitions and one copy of the text. The \emph{gather\_kv} mode all-gathers visual K/V and combines them with local text K/V. This gives each rank the full key set, but each logical core must then hold the whole visual K/V sequence in its 24\,GiB of HBM, so per-rank K/V memory does not shrink as the CP group grows. Ulysses and Ring organize these inputs differently, so their masks must follow the key order of each kernel call.

\subsection{Representing Attention Mask}
\label{sec:parallel-attention-mask}

Neuron compiles attention calls for fixed tensor shapes, so shorter text inputs retain padding. Padding keys must be excluded from softmax to preserve the weights on valid keys. When valid keys form a contiguous interval, \name passes its bounds to AWS's fused attention kernel, which combines score computation, softmax, and value aggregation~\cite{aws-nki-attention-cte,aws-nki-attention-source}. For each tile of 128 queries, the Tensor Engine computes $QK^\top$ into PSUM. As the Vector Engine moves the scores to SBUF, it replaces those outside the interval with $-\infty$ and tracks each row's running maximum; the Scalar Engine then computes the exponentials and their row sums. After a DMA transpose, the Tensor Engine multiplies the probabilities by $V$, so the score matrix never returns to HBM. The kernel reads two bounds per query from HBM instead of a mask entry per query--key pair.

For HunyuanVideo, \name arranges the full joint key sequence as visual tokens, valid text tokens, and text padding. Let $N_{\mathrm{visual}}$ be the visual sequence length and $L_{\mathrm{text}}$ the valid text length obtained from the input mask. Every query $i$ uses the allowed interval $[\ell_i,u_i)$, where
\begin{equation}
    \ell_i = 0, \qquad u_i = N_{\mathrm{visual}} + L_{\mathrm{text}}.
    \label{eq:parallel-joint-bounds}
\end{equation}
Varying $L_{\mathrm{text}}$ changes the allowed keys while preserving the compiled tensor shapes, and padding never enters the attention weights or their normalization.

\subsection{Distributed Joint Attention}
\label{sec:distributed-joint-attention}

\begin{figure}[t]
    \centering
    \includegraphics[width=\columnwidth]{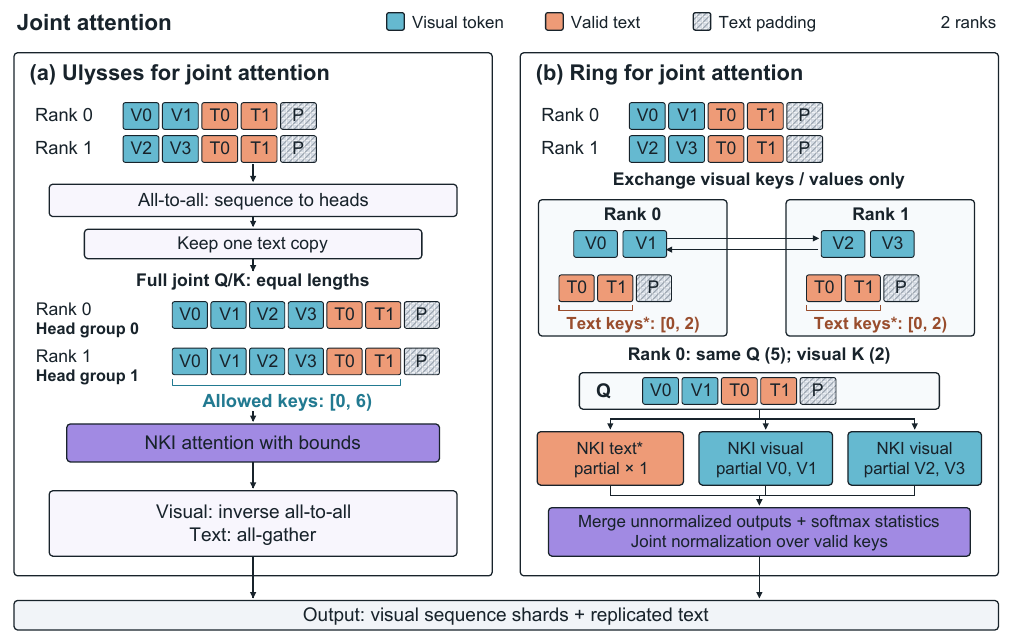}
    \caption{Joint attention with visual tokens partitioned and text replicated across two ranks. (a) Ulysses gives each head group the full joint Q/K sequence with equal lengths; bounds $[0,6)$ exclude text padding. (b) Ring uses the same local queries in separate text and visual calls with unequal Q/K lengths, then merges unnormalized outputs and softmax statistics. Starred: the text bounds $[0,2)$ that Ring needs for padded prompts. Both paths restore visual sequence shards and replicated text.}
    \label{fig:parallel-joint-attention}
\end{figure}

Figure~\ref{fig:parallel-joint-attention} compares two ways to provide the joint key set to the attention computation. Ulysses partitions attention heads across ranks, so each rank processes the full joint sequence for its assigned heads. Ring keeps queries local and evaluates them against successive K/V partitions. Both return partitioned visual outputs and replicated text outputs, as required by the next layer.

\textbf{Ulysses for joint attention.} In \emph{gather\_kv}, queries cover the local visual partition and all text, while keys cover the full visual and text sequences, so query and key lengths differ. \name instead uses Ulysses to exchange sequence partitions for head partitions~\cite{ulysses}. Each rank then holds the full joint sequence for its assigned heads, with equal Q and K lengths.

An all-to-all gives each logical core the full visual Q/K/V sequence for $H/(\mathrm{TP}\cdot\mathrm{CP})$ of the model's $H$ heads. Applying the same exchange to replicated text delivers identical copies of that rank's text head partition. Each rank retains one copy so that text keys enter softmax only once. It then concatenates visual and text Q/K/V and runs fused attention, using the bounds in Equation~\ref{eq:parallel-joint-bounds} for padded text.

The next layer expects the original token distribution. An inverse all-to-all therefore returns visual outputs to sequence partitions, while an all-gather across head partitions restores replicated text outputs.

\textbf{Ring for joint attention.} Ring attention keeps queries local while rotating K/V partitions among ranks~\cite{ringattention}. Here Q contains local visual tokens and all text, whereas a rotating K/V partition contains only visual tokens. The AWS Ring kernel used here requires equal query and key lengths~\cite{aws-nki-ring-source}. \name therefore builds the ring from separate fused attention calls that accept unequal lengths without a mask.

Each call uses the same local Q with either text K/V or a visual K/V partition. Text is processed once per rank, so its keys are counted once; only visual K/V circulate. Each rank obtains one text partial and one partial for each visual partition, including its own.

Independently normalized partials cannot be added, because each uses a different softmax denominator. For each query, each call therefore returns an unnormalized weighted value sum, the maximum score, and the sum of exponentials relative to that maximum. The kernel keeps these statistics in SBUF and writes them to HBM in its tile layout, one query per partition, which \name reorders to query order. Outside the kernel, \name rescales both sums to a common maximum and adds them across partitions. Dividing the combined weighted sum by the combined exponential sum restores joint normalization without gathering all visual K/V~\cite{flashattention,ringattention}. The outputs retain Q's layout: local visual tokens and replicated text.

The text partial contains only text keys, so masking its padding requires bounds $[0,L_{\mathrm{text}})$ in local text-key coordinates. Padding must be excluded from both its weighted value sum and softmax statistics before the joint merge.

\subsection{Other Parallel Modes}
\label{sec:parallel-modes-support}

TP, SP, and CP divide work within one denoiser pass. \textbf{Classifier-free guidance (CFG) parallelism} executes the conditional and unconditional passes concurrently in separate rank groups~\cite{cfg,xdit}. \textbf{Data parallelism (DP)} assigns independent requests to complete model replicas, increasing throughput. Each replica can combine TP with CP or CFG parallelism.

Ulysses partitions each TP rank's heads across the CP group, so the total head count must be divisible by the product of the TP and CP group sizes, which count logical cores. Table~\ref{tab:parallel-execution-support} in Section~\ref{subsec:parallelism-eval} lists the modes each model runs.

\section{Adaptive Step Caching on Trainium}
\label{sec:caching}

Nearby diffusion steps can produce similar denoiser outputs. \name uses this similarity to replace some full denoiser executions with estimates. The skip decision changes from step to step, so \name makes it on the host, outside the compiled denoiser. The host stores the previous output used by the generation loop and its most recently measured change. A full execution measures the change relative to the previously used output, then saves the new output. A cached step adds the saved change to the previous output and stores the estimate for the next step. Both paths apply the same numerical update rule to the latent, reducing denoiser calls while retaining every sampling step. This output estimator supports three caching modes: \emph{fixed cadence}, \emph{calibrated adaptive}, and \emph{online-delta}. They differ in when to replace a full denoiser execution with an estimate.

\begin{figure}[t]
    \centering
    \includegraphics[width=\columnwidth]{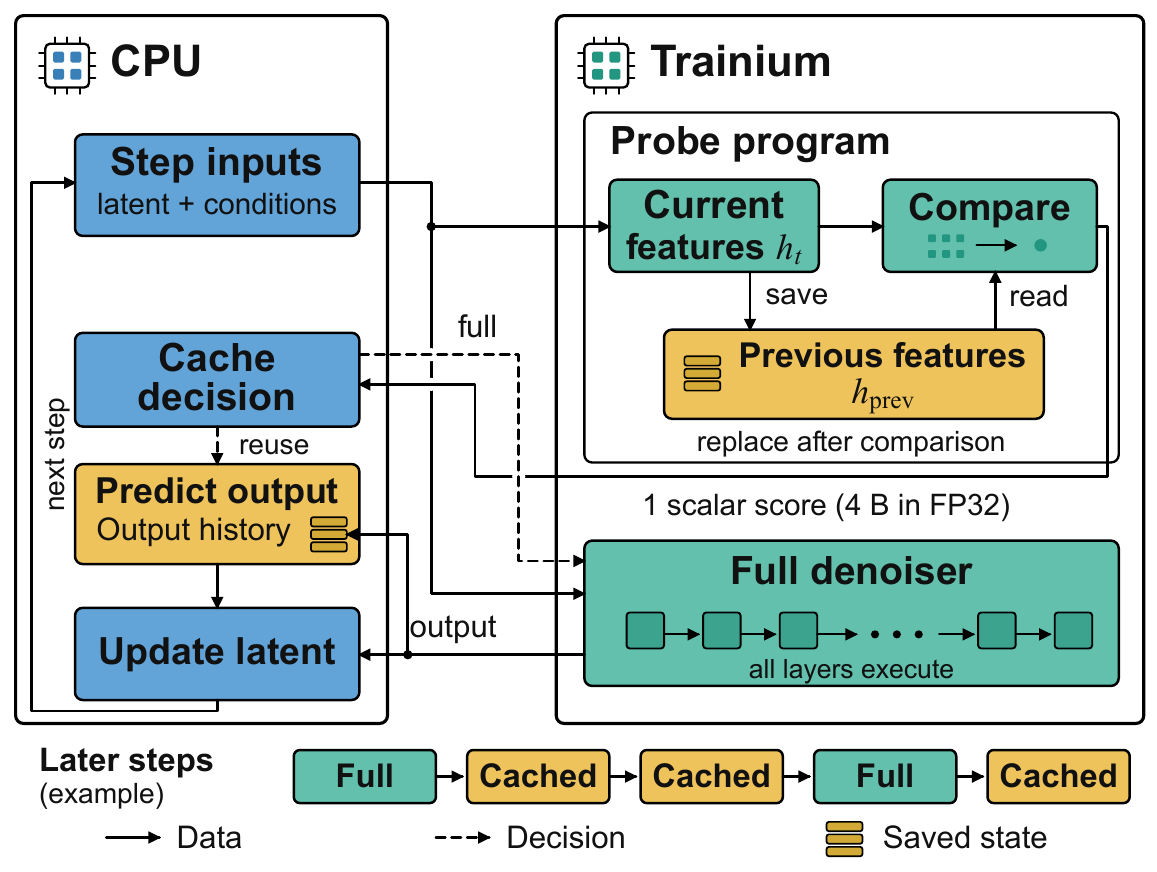}
    \caption{Calibrated adaptive caching with a device probe in \name. The probe retains features on Trainium and returns a scalar score to the host CPU. Teal denotes full denoiser execution; gold denotes cached prediction. Solid arrows show data flow; dashed arrows show execution choices.}
    \label{fig:caching-system-trainium}
\end{figure}

The calibrated adaptive mode needs a signal that costs less to obtain than the computation it can save. TeaCache~\cite{teacache} motivates using changes in early model features to estimate later output changes. On Trainium, these features remain inside a program compiled before inference. Observing them requires the program to expose a result, but returning a large feature array to the host for comparison would introduce communication overhead. \name compiles a shallow computation, called a probe, to produce the change score directly on Trainium. Figure~\ref{fig:caching-system-trainium} shows how this path connects device computation and saved features to the host's reuse decision.

The device probe executes the denoiser's input projection and the first block's normalization and modulation, using the current timestep and conditioning inputs. It stops before attention. The compiler removes the unused later computations. Wan, HunyuanVideo, and LTX-2 load the result as a separate program; FLUX and Qwen-Image fuse it into the denoiser program. Full steps still run the complete denoiser, so the probe never changes what a full step computes.

The probe compares its current features with those saved from its previous call. In Figure~\ref{fig:caching-system-trainium}, these arrays are labeled \emph{Current features} ($h_t$) and \emph{Previous features} ($h_{\mathrm{prev}}$). If $D$ measures their difference, one call computes a scalar change score and updates the saved features:
\begin{equation}
    d_t = D(h_t,h_{\mathrm{prev}}), \qquad
    h_{\mathrm{prev}} \leftarrow h_t.
    \label{eq:cache-probe-state}
\end{equation}
The score $d_t$ is computed before the saved array is replaced. For example, a probe can divide the average absolute change by the average magnitude of the saved features. Other measures retain the same scalar interface.

To keep the saved features on Trainium across calls, \name assigns the probe's new feature output to their existing memory through input--output aliasing~\cite{aws-neuronx-trace}. After comparison, the current features replace the old ones there, and the next call reads them directly. Feature computation, comparison, and state update thus occur within one probe call. Only the scalar score returns to the host, reducing probe output transfer from a feature array to a single value. The saved features support the next comparison; the separate host output history supplies cached predictions.

In calibrated adaptive mode, a host controller uses the score to choose full execution or prediction. Before inference, reference runs provide examples of feature changes and corresponding output changes. This calibration fits a curve relating the two. The controller compares the estimated output change with a threshold, either separately for each step or accumulated across steps. Reaching the threshold triggers full execution and resets the sum when accumulation is used. All three modes retain initial full steps to establish output history and final full steps to limit reuse near completion. These decisions remain outside the compiled denoiser.

\begin{figure*}[t]
    \centering
    \includegraphics[width=\textwidth]{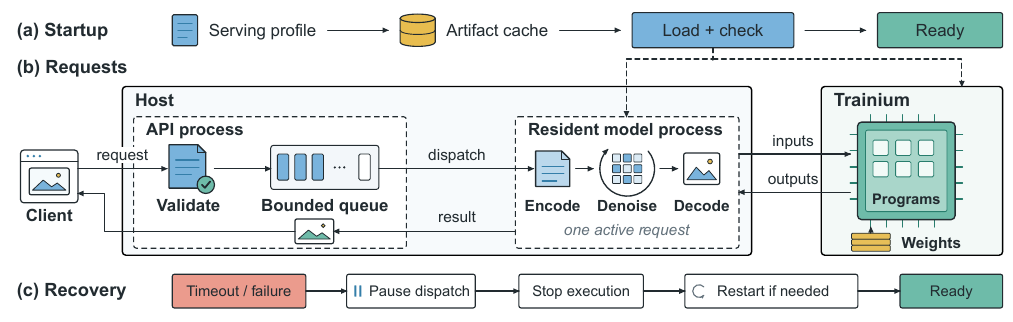}
    \caption{Serving in \name. (a) Startup selects artifacts for the serving profile, loads components, and checks readiness. Dashed arrows connect loading to the model process and Trainium. (b) The API process validates and queues requests. One resident model process orchestrates encoding, denoising, and decoding and controls the device cores; component placement depends on the model configuration. (c) After a timeout or failure, dispatch waits for prior execution to stop and the model process to become ready, with a restart when needed.}
    \label{fig:serving-arch}
\end{figure*}

Online-delta uses outputs already available on the host, avoiding a separate probe call. At a full step, it measures output change relative to the previously used output, which may have been estimated. If this change is below $\alpha$ times the first change measured in the generation, the next step is cached. That cached step must be followed by full execution to obtain a fresh measurement. Calibration can select this mode when early features poorly predict output changes but output changes remain related across neighboring steps. If neither signal is suitable, it selects fixed cadence, which caches at a preset interval without a change signal. Both modes are probe-free; their effect on final quality must be assessed through complete image or video generation.

Because a separate probe loads only the weights of the layers it runs, it fits beside the denoiser even on HunyuanVideo. Section~\ref{sec:cache-effectiveness} measures what the probe costs per step and how well its signal predicts the output change.

\section{Serving Layer}

\label{sec:serving}

Serving applies the same boundary across requests: programs and weights stay loaded, while each request's sampling and caching state starts fresh on the host. \name separates request handling from model execution using two host processes (Figure~\ref{fig:serving-arch}). The \emph{API process} validates requests, manages the queue, and returns results. A single \emph{resident model process} loads the components for text encoding, denoising, and decoding, and retains them across requests. The process coordinates all three stages and manages the assigned cores. Keeping them resident avoids repeated runtime initialization and weight loading. This separation allows the model process to be restarted without restarting the API process.

Components that stay resident together share the same cores, which constrains both their parallel configurations and their memory. In our NxD execution path, resident components share the same set of cores, although their TP degrees can differ. For FLUX, the denoiser's weights are partitioned across four cores (TP=4), while each core holds a full copy of the VAE weights (TP=1). Each component must be compiled for this shared set of cores and its own TP degree. Device memory must accommodate all resident weights, including the VAE replicas, and execution buffers.

At startup, \name configures the model process using a serving profile that specifies the model, supported input shapes, data type, and parallel configuration. It selects the corresponding compiled artifacts and checks that their core assignments match this configuration before loading them. The model process then runs a test generation to check that the loaded components work together. It validates the output before becoming ready to execute requests (Figure~\ref{fig:serving-arch}(a)). After a restart, the process reloads the saved artifacts and repeats these checks.

During serving, the API process checks requests against the serving profile and places accepted requests in a bounded queue, rejecting new requests when the queue is full. The model process handles queued requests one at a time to prevent concurrent changes to the pipeline's sampling and caching state. Each request needs its own sampling and caching state. The model process therefore initializes the latent and timestep schedule for each generation. When step caching is enabled, it also clears the controller's output history to prevent reuse of predictions from a previous request. The model process returns the result to the API process (Figure~\ref{fig:serving-arch}(b)).

\noindent\textbf{Recovery.} When a request times out or the model process fails, \name pauses dispatch before attempting recovery. A timeout may leave device computation running, so \name first requests cancellation and waits for the current generation to stop. If cancellation succeeds, the same model process can handle subsequent requests. Otherwise, \name terminates the process and confirms its exit before starting a replacement. A model process failure also triggers replacement. The new process reloads the saved artifacts and passes the startup checks before dispatch resumes (Figure~\ref{fig:serving-arch}(c)). If recovery fails, dispatch remains paused.

\section{Evaluation}
\label{sec:evaluation}

We evaluate \name on five image and video models and ask five questions: how fast \name is end to end (\S\ref{sec:eval-e2e}), which mechanism the speedup comes from (\S\ref{sec:eval-attribution}), which parallel configurations pay off (\S\ref{subsec:parallelism-eval}), what step caching saves and costs (\S\ref{sec:cache-effectiveness}), and what resident serving buys in startup and recovery (\S\ref{sec:serving-eval}).

\subsection{Experimental Setup}
\label{sec:eval-setup}

\paragraph{Platform.}
All experiments run on one AWS \texttt{trn2.3xlarge} instance~\cite{aws-trn2-instances} with one Trainium2 chip, 96\,GiB of HBM~\cite{aws-trn2-arch}, and 124\,GiB of usable host memory. Under LNC=2 the chip exposes four logical cores, each pairing two physical NeuronCores~\cite{aws-lnc}. All experiments use one locked software stack: torch 2.9.1, torch-neuronx 2.9.0, neuronx-cc 2.26, neuronx-distributed 0.19, and NKI 0.5. Unless stated otherwise, the denoiser runs in BF16 with four-way tensor parallelism (TP=4) and step caching off.

\paragraph{Workloads.}
Table~\ref{tab:workloads} lists the five models. Step count, guidance, and negative prompt follow each model's official defaults except where Table~\ref{tab:workloads} notes otherwise, and every comparison within a model fixes the checkpoint revision, prompt, seed (42), and output shape. Wan~2.1 runs at its official 81 frames, 32.8k visual tokens per step; its VAE decodes on the device in FP32, in fixed-size temporal chunks, and agrees with the diffusers FP32 decoder to a maximum absolute difference of $5.6{\times}10^{-6}$ (relative L2 error $1.2{\times}10^{-6}$, compared over the first 33 frames).

\begin{table}[t]
\caption{Evaluated models. Tokens are the visual tokens the denoiser attends over per step; guidance is distilled (dist.), classifier-free (CFG), or embedded (emb.).}
\label{tab:workloads}
\centering
\footnotesize
\setlength{\tabcolsep}{2.2pt}
\resizebox{\columnwidth}{!}{%
\begin{tabular}{@{}lrrrrl@{}}
\toprule
model & params & H$\times$W$\times$F & tokens & steps & guidance \\
\midrule
FLUX.1-dev   & 12B & $1024^2$                      & 4{,}096  & 50 & 3.5 dist. \\
Qwen-Image   & 20B & $1024^2$\,$^{b}$              & 4{,}096  & 50 & 4.0 CFG \\
Wan~2.1      & 14B & $480{\times}832{\times}81$        & 32{,}760 & 50 & 5.0 CFG \\
HunyuanVideo & 13B & $320{\times}512{\times}61$\,$^{a}$ & 10{,}240 & 50 & 6.0 emb. \\
LTX-2        & 19B & $480{\times}704{\times}49$\,$^{a}$ & 2{,}310  & 40 & 1.0\,$^{c}$ \\
\bottomrule
\end{tabular}}
\par\smallskip
{\raggedright\scriptsize
$^{a}$Below the official shape (HunyuanVideo $720\times1280\times129$, LTX-2 $512\times768\times121$), which does not fit on one chip; LTX-2 runs the first stage of its two-stage pipeline. $^{b}$The official 1:1 shape is $1328\times1328$; both image models run at $1024^2$ so that they process the same number of tokens. $^{c}$The official first stage uses CFG 4.0; \name does not yet build the batch-2 program CFG needs, so LTX-2 runs without guidance.\par}
\vskip -0.1in
\end{table}

\paragraph{Baselines.}
Table~\ref{tab:baselines} lists the systems we compare and the mechanisms each one has. \emph{PyTorch/XLA} is \name with every NKI attention call replaced by PyTorch's scaled dot-product attention (SDPA) compiled through XLA~\cite{pytorch-xla}. The rest of the pipeline is unchanged---the same programs for the other components, presharded weights, and step caching---so the difference isolates the attention kernel. Because \texttt{difflet serve} does not accept SDPA, this baseline runs standalone only, and the ring mode of context parallelism has no SDPA variant. \emph{NxDI}~\cite{nxdi} is the FLUX.1-dev application AWS provides for Trainium. It compiles the pipeline ahead of time with the same family of NKI attention kernels, but shards the checkpoint at load time in every process and supports tensor parallelism only; its public release has no pipeline for the other four models. \name is the only system that combines all six mechanisms, and Section~\ref{sec:eval-attribution} measures what each contributes.

\begin{table}[t]
\caption{Systems compared and the mechanisms each provides: ahead-of-time compilation, NKI attention, presharded weights, resident serving, step caching, and parallelism. NxDI covers FLUX.1-dev only. Y: yes; $\times$: no.}
\label{tab:baselines}
\centering
\footnotesize
\setlength{\tabcolsep}{2.4pt}
\resizebox{\columnwidth}{!}{%
\begin{tabular}{@{}lcccccl@{}}
\toprule
system & AOT & NKI & preshard & resident & cache & parallel \\
\midrule
PyTorch/XLA & Y & $\times$ & Y        & $\times$$^{d}$ & Y        & TP, CP$^{e}$ \\
NxDI        & Y & Y        & $\times$ & $\times$       & $\times$ & TP \\
\name       & Y & Y        & Y        & Y              & Y        & 5 axes$^{f}$ \\
\bottomrule
\end{tabular}}
\par\smallskip
{\raggedright\scriptsize $^{d}$\texttt{difflet serve} does not accept SDPA. $^{e}$No ring mode: it has no SDPA variant. $^{f}$Tensor, sequence, context (gather, Ulysses, ring), CFG, and data parallelism.\par}
\vskip -0.1in
\end{table}

\paragraph{Measurements.}
\emph{Request latency} is the time to generate one output with the model already resident, measured through \texttt{difflet serve}, over HTTP for Wan and LTX-2, where returning the output adds well under a second. \emph{Standalone latency} additionally covers process start and model loading: a \emph{cold} run first flushes the host page cache, a \emph{warm} run reuses the cached weight files but still loads them onto the device. \emph{Denoising time} is the wall time of the denoising loop, including steps served from the cache, and \emph{step latency} is the median time of an uncached step, excluding the first; with sequential CFG it covers both passes. Compilation is a one-time build into empty caches and is reported separately. Resident request latency is the median of five requests after one warm-up for image models and of three for video models; standalone runs and compilations are measured once. The device slows by up to 25--30\% under back-to-back load and recovers after 90\,s of idle, so every resident measurement idles the device for 90\,s first.

\paragraph{Quality.}
Step caching is judged against the uncached output of the same model, prompt, and seed by PSNR~\cite{wang2009mse} and LPIPS~\cite{lpips}, frame-averaged for video, over 16 prompts per modality: eight from DrawBench~\cite{drawbench-imagen} and eight from PartiPrompts~\cite{partiprompts} for images, 16 from VBench~\cite{vbench} for video, none used for calibration.\footnote{$\mathrm{PSNR} = 10 \log_{10}(255^2/\mathrm{MSE})$ in dB over 8-bit pixels; every 10\,dB less is a tenfold larger mean squared difference.} Wan, at about 50 minutes per prompt for its four configurations, is scored by PSNR only, on five VBench prompts: the first five of the overall-consistency list. We report the mean and the worst prompt.

\subsection{End-to-End Performance}
\label{sec:eval-e2e}

Table~\ref{tab:e2e} reports, for each model, the one-time compilation, the standalone latency of a fresh process, the request latency of a resident one, and the step latency with each attention implementation.

\begin{table}[t]
\caption{End-to-end latency at TP=4. Cold and warm are standalone runs; resident is the request latency of the model process; step is the step latency with NKI and with SDPA attention.}
\label{tab:e2e}
\centering
\footnotesize
\setlength{\tabcolsep}{2.2pt}
\resizebox{\columnwidth}{!}{%
\begin{tabular}{@{}lrrrrrr@{}}
\toprule
& compile & \multicolumn{3}{c}{latency (s)} & \multicolumn{2}{c}{step (ms)} \\
\cmidrule(lr){3-5}\cmidrule(lr){6-7}
model & (min) & cold & warm & resident & NKI & SDPA \\
\midrule
FLUX.1-dev   & 16.7  & 318   & 47.6  & 14.2  & 270     & 682 \\
\quad NxDI   & 15.7  & 337   & 66.3  & --    & 271     & -- \\
Qwen-Image   & 24.1  & 537   & 98.9  & 42.4  & 835     & 1{,}582 \\
Wan~2.1      & 75.1  & 1{,}311 & 974 & 1{,}245 & 14{,}876 & --$^{g}$ \\
HunyuanVideo & 85.8  & 625   & 141   & 59.9  & 845     & 3{,}620 \\
LTX-2        & 17.6  & 780   & 64.7  & 43.3  & 414     & 514 \\
\bottomrule
\end{tabular}}
\par\smallskip
{\raggedright\scriptsize $^{g}$The SDPA program does not compile at 81 frames: it needs 7.96M instructions against the compiler's limit of 5M.\par}
\vskip -0.1in
\end{table}

\paragraph{Findings.}
\emph{The attention kernel speeds up every denoiser step, and resident serving removes most of what a standalone run adds around it.} Against PyTorch/XLA's SDPA, \name's denoiser step is 1.24--4.28$\times$ faster on the four models whose SDPA program compiles, and at Wan's 81 frames only the fused kernel compiles at all. The model process then serves a request 1.49--3.35$\times$ faster than a warm standalone run on FLUX, Qwen-Image, HunyuanVideo, and LTX-2; Wan is the exception, for a reason in its decoder. Section~\ref{sec:eval-attribution} explains it and splits the gain from PyTorch/XLA to \name by mechanism.

\emph{Once a model is resident, request latency is mostly denoising---except on the video pipelines with heavy stages around the denoiser.} Denoising takes 95\% of a FLUX request and 98\% of a Qwen-Image request, so these are the models that caching and parallel execution speed up most directly. Wan spends about 42\% of its resident request (520\,s) in the VAE decode, which the model process runs on the host (Section~\ref{sec:eval-attribution}), HunyuanVideo 30\% in its encoders and decoder, and LTX-2 three fifths (27 of 43\,s) in the stages around its denoiser.

\emph{\name's generality costs no denoising performance against NxDI.} On FLUX.1-dev, the one model NxDI supports, both systems run a 270\,ms denoiser step, because both use the same family of NKI attention kernels. \name is faster end to end because of loading: NxDI loads the full diffusers pipeline on the host in every process and warms up each component, whereas \name reads checkpoints already split per core at compile time. Warm latency falls from 66.3 to 47.6\,s and the Neuron weight load from 43 to 23\,s, for a compilation one minute longer (16.7 against 15.7\,min), into which the checkpoint splitting moves.

\subsection{Where the Speedup Comes From}
\label{sec:eval-attribution}

\begin{figure}[t]
\centering
\includegraphics[width=\columnwidth]{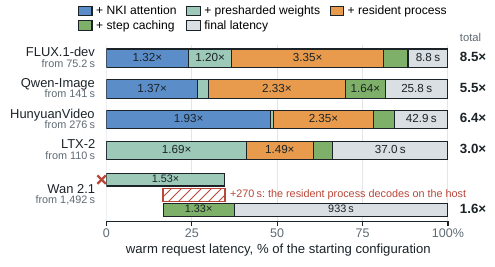}
\caption{Where the request-latency speedup comes from, adding \name's mechanisms one at a time to a standalone PyTorch/XLA run. Each bar is a model's starting latency; colored segments are the time each mechanism removes, labeled with that step's speedup where it fits, and gray is the final latency. $\times$: Wan's PyTorch/XLA program does not compile, so its bar starts from NKI attention. Caching is fixed cadence~2; for FLUX, Wan, and LTX-2 it is the resident caching-off request minus the denoising time cadence~2 saves. LTX-2's first two stages are medians of two alternated runs (106--113\,s).}
\label{fig:attribution}
\end{figure}

Figure~\ref{fig:attribution} starts from a standalone PyTorch/XLA run and adds \name's mechanisms one at a time. To separate the attention kernel from the loader, the first two stages shard the checkpoint at load time, as a system without \name's compile-time splitting would.

\paragraph{Findings.}
\emph{Each mechanism dominates on different models.} Together they cut request latency by 3.0--8.5$\times$ on the four models with a PyTorch/XLA baseline. Resident serving is the largest single step on the image models (3.35$\times$ on FLUX, 2.33$\times$ on Qwen-Image) and on HunyuanVideo (2.35$\times$), because a standalone request pays process start, pipeline construction, and a 10--140\,s device weight load on every call. The attention kernel gives 1.93$\times$ on HunyuanVideo and 1.32--1.37$\times$ on the image models; on LTX-2 its 1.24$\times$ faster denoiser step would save 4\,s of a 110\,s run, and the measured difference (0.1\,s) is within the run-to-run spread (106--113\,s). Presharding matters where the checkpoint is large relative to the rest of the pipeline: 1.69$\times$ on LTX-2, whose 19B denoiser takes 38--42\,s to shard on the host, and 1.53$\times$ on Wan, whose text encoder and denoiser take 173 and 354\,s. Step caching adds 1.40--1.64$\times$ where the request is mostly denoising, 1.33$\times$ on Wan, and 1.17$\times$ on LTX-2.

\emph{Wan is the exception to resident serving, because its model process decodes on the host.} Its resident request takes 1{,}245\,s against 974\,s standalone. Both paths run the same denoiser program and denoise in about 720\,s; the standalone run then decodes on the device with the chunked FP32 VAE in 126\,s, whereas the model process decodes on the host in about 520\,s, more than the 140\,s weight load it saves. The chunked decoder does not fit beside the resident model: the 81-frame denoiser and the text encoder already take about 20\,GiB of each logical core's 24\,GiB, and the decoder's largest chunk moves 3\,GiB of state in and 3\,GiB out.

\begin{figure}[t]
\centering
\includegraphics[width=\columnwidth]{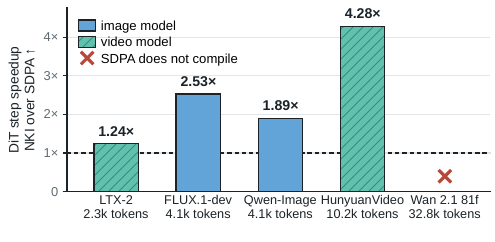}
\caption{Step speedup of NKI attention over PyTorch/XLA's SDPA, each model at its evaluated shape, ordered by visual tokens per step. $\times$: Wan~2.1's SDPA program does not compile at 81 frames.}
\label{fig:attn-tokens}
\end{figure}

\emph{The kernel's gain grows with sequence length, and at 32.8k tokens it is the only attention that runs.} Attention cost grows quadratically with the number of tokens. Figure~\ref{fig:attn-tokens} orders the models by sequence length: the step speedup rises from 1.24$\times$ on LTX-2 (2.3k tokens) through 1.89--2.53$\times$ on the image models (4.1k) to 4.28$\times$ on HunyuanVideo (10.2k). At Wan's 32.8k tokens the SDPA program does not compile, so the fused kernel is the condition for running at all. The models differ in architecture as well as length, so the trend is across workloads rather than a controlled sweep.

\subsection{Parallel Execution}
\label{subsec:parallelism-eval}

\begin{figure}[t]
\centering
\includegraphics[width=\columnwidth]{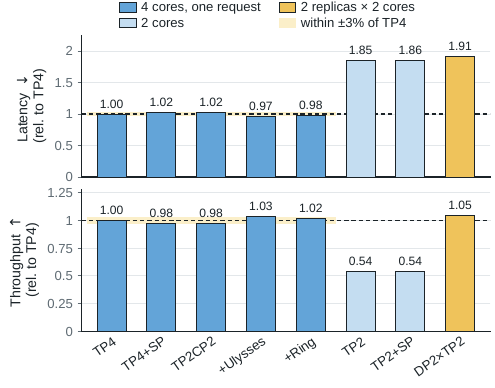}
\caption{Parallel configurations for FLUX.1-dev (4.1k tokens) relative to TP4: resident request latency and throughput. DP2$\times$TP2 serves two concurrent requests; its latency is the wall time for both and its throughput counts both.}
\label{fig:parallel}
\end{figure}

We sweep every configuration \name can launch on the four cores for FLUX.1-dev (4.1k tokens) in the model process: tensor parallelism alone (TP4, TP2), with Megatron-style sequence parallelism (SP), with context parallelism over two cores in each of its three modes (TP2CP2), and two TP2 replicas serving concurrent requests (DP2$\times$TP2). Each configuration is compiled separately and measured as in Section~\ref{sec:eval-setup}. For the long sequence, we attempt Wan~2.1 at 81 frames (32.8k tokens) in the configurations that change the most at that length: sequence parallelism (TP4+SP), Ulysses context parallelism (TP2CP2), TP2 with CFG parallelism, which runs the two guidance branches on separate core pairs, and the data-parallel pair. Figure~\ref{fig:parallel} normalizes the FLUX results to TP4, and Table~\ref{tab:parallel} gives the absolute latencies for both models; Table~\ref{tab:parallel-execution-support} lists the parallel modes each model runs.

\begin{table}[t]
    \centering
    \caption{Parallel modes each model runs. Y: supported; $\dagger$: constrained; $\times$: unsupported; --: not applicable.}
    \label{tab:parallel-execution-support}
    {\footnotesize
    \setlength{\tabcolsep}{2pt}
    \begin{tabular}{@{}lccccccc@{}}
        \toprule
        & & \multicolumn{3}{c}{CP} & & & \\
        \cmidrule(lr){3-5}
        Model & TP & Gather & Ring & Ulysses & SP & CFG & DP \\
        \midrule
        FLUX.1-dev   & Y & Y & $\dagger^{1}$ & Y & Y & -- & Y \\
        Qwen-Image   & Y & Y & Y & Y & Y & $\times$ & Y \\
        Wan 2.1/2.2  & Y & Y & $\dagger^{1}$ & $\dagger^{4}$ & Y & $\dagger^{4}$ & $\dagger^{4}$ \\
        HunyuanVideo & Y & $\dagger^{2}$ & $\times^{3}$ & Y & Y & -- & Y \\
        LTX-2        & Y & $\times$ & $\times$ & $\times$ & $\times$ & Y & Y \\
        \bottomrule
    \end{tabular}
    \par\smallskip
    \raggedright
    Gather denotes \emph{gather\_kv}. Context parallelism is not combined with SP or CFG parallelism. $^{1}$The ring kernel used by FLUX and Wan requires per-rank sequence lengths that are multiples of 128 tokens, because it processes queries in tiles of 128, one per SBUF partition. $^{2}$HunyuanVideo's masked \emph{gather\_kv} path failed to compile at CP size two with Neuron compiler 2.26. $^{3}$Ring does not yet mask text padding (Section~\ref{sec:distributed-joint-attention}). $^{4}$Not at Wan~2.1's 81 frames on one chip: Ulysses (TP2CP2) needs 43\,MiB more HBM than is free, and the TP2-based CFG and DP programs exhaust the compiler's host memory (Table~\ref{tab:parallel}).
    \par
    }
\end{table}

\begin{table}[t]
\caption{Request latency and step latency across parallel configurations. FLUX is resident; Wan is standalone. DP rows serve two concurrent requests, so their latency is the wall time for both. FLUX TP4 is a separate run from Table~\ref{tab:e2e} (14.2\,s, 270\,ms).}
\label{tab:parallel}
\centering
\footnotesize
\setlength{\tabcolsep}{4pt}
\resizebox{\columnwidth}{!}{%
\begin{tabular}{@{}lrrrr@{}}
\toprule
& \multicolumn{2}{c}{FLUX (4.1k tok.)} & \multicolumn{2}{c}{Wan 81f (32.8k tok.)} \\
\cmidrule(lr){2-3}\cmidrule(lr){4-5}
config & req.\ (s) & step (ms) & req.\ (s) & step (ms) \\
\midrule
TP4               & 13.97 & 272.5 & 974 & 14{,}876 \\
TP4+SP            & 14.28 & 278.7 & 933 & 13{,}882 \\
TP2CP2, gather    & 14.29 & 278.8 & --  & --  \\
TP2CP2, Ulysses   & 13.51 & 263.3 & HBM$^{q}$ & HBM$^{q}$ \\
TP2CP2, ring      & 13.75 & 268.2 & --  & --  \\
TP2               & 25.83 & 510.3 & OOM$^{k}$ & OOM$^{k}$ \\
TP2+SP            & 25.97 & 511.3 & --  & --  \\
TP2+CFG           & n/a$^{j}$ & n/a$^{j}$ & OOM$^{k}$ & OOM$^{k}$ \\
\midrule
DP2$\times$TP2    & 26.68 & 529.6 & OOM$^{k}$ & OOM$^{k}$ \\
\bottomrule
\end{tabular}}
\par\smallskip
{\raggedright\scriptsize --: not attempted. $^{j}$FLUX is guidance-distilled and has no second guidance branch. $^{k}$Does not compile: the compiler backend exhausts the host's 124\,GiB of memory; DP2$\times$TP2 replicas run the TP2 program. $^{q}$Compiles but does not load: the static DMA rings for its spill traffic need 43\,MiB more HBM than is free beside the resident text encoder.\par}
\vskip -0.1in
\end{table}

\paragraph{Findings.}
\emph{At 4.1k tokens, tensor parallelism over all four cores is the latency default.} Every configuration that keeps four cores on one request stays within 3.5\% of TP4 (Table~\ref{tab:parallel}). Sequence parallelism and gather-based context parallelism add 2\% per step (278.7 and 278.8 against 272.5\,ms); Ulysses and ring attention remove 3\% and 2\% (263.3 and 268.2\,ms), half a second on a 14\,s request. At this length the activations that sequence parallelism shards and the keys and values that context parallelism exchanges are small, so the extra collectives cost about what the split saves. Ulysses, which trades the sequence split for a head split with one all-to-all on each side of attention, is the fastest of the three context-parallel modes; it and ring attention are the only configurations faster than TP4, by 3\% and 2\%. Halving the cores to TP2 costs 1.87$\times$ per step, so tensor parallelism scales from two to four cores at 94\% efficiency.

\emph{Data parallelism buys throughput only marginally once models are resident.} Two TP2 replicas, started together on separate core pairs, finish two requests in 26.7\,s, against 27.9\,s for TP4 serving them in turn: 5\% more throughput for almost twice the latency. Each replica's step (529.6\,ms) is 4\% slower than a TP2 replica running alone, so the replicas contend slightly for the chip's shared resources. Data parallelism therefore pays only when per-request work outside the device, such as loading, dominates; in a resident server it does not.

\emph{At 32.8k tokens, sequence parallelism pays, and the configurations that split the model in halves do not run.} Sequence parallelism splits across the four cores the per-token work that tensor parallelism repeats on each of them over the full sequence---the norms, modulation, and residual additions---and at this length that cuts Wan's denoiser step by 7\% (13.9 against 14.9\,s) and its warm request by 4\% (933 against 974\,s). Ulysses context parallelism over two TP2 pairs compiles but does not load: the static DMA rings the compiler emits for its spill traffic need 43\,MiB more HBM than is free beside the resident text encoder. The TP2 program, with or without CFG batching, holds half the model and the full 32.8k-token sequence on each core, and the compiler backend exhausts the host's 124\,GiB building it, in both attempts. CFG parallelism, which at short sequences removes work rather than moving it, and the data-parallel pair are therefore unavailable at this length, and the limit is the compiler's host memory rather than the device's.

\emph{Ulysses context parallelism handles padded prompts correctly, to within BF16 rounding.} Under Ulysses, each core holds a slice of the sequence and the all-to-all hands it a slice of the heads over the whole sequence, so the text padding mask travels as a count of valid keys. We run the HunyuanVideo denoiser (two dual-stream and two single-stream blocks at $320{\times}512{\times}61$) as TP2CP2 with Ulysses and as TP2 alone, one compiled program per configuration, at five prompt lengths from 5 to 256 valid text tokens passed as a runtime input. The two configurations differ by at most $7.8{\times}10^{-2}$ (mean $3.7{\times}10^{-3}$, cosine similarity 0.99992) on outputs of magnitude up to 3.3, a few BF16 rounding steps, and the difference does not grow with prompt length. Changing the prompt length moves the output 8--27$\times$ more than that (mean $3.1{\times}10^{-2}$ to $1.0{\times}10^{-1}$), so a mask the split mishandled would show.

\subsection{Step Caching}
\label{sec:cache-effectiveness}

\begin{figure}[t]
\centering
\includegraphics[width=\columnwidth]{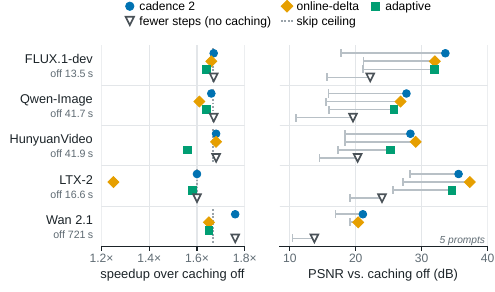}
\caption{Step caching at official step counts. Left: denoising-time speedup over caching off (caching-off time under each model); dotted: the ceiling set by cadence~2's 20 of 50 skipped calls (15 of 40 on LTX-2); fewer steps runs cadence~2's 30 calls (25 on LTX-2) without caching. Right: PSNR against the model's caching-off output, mean over 16 prompts (five on Wan), whisker to the worst prompt. Online-delta's speedup is for the timing prompt. Calibrated adaptive runs standalone, against the standalone caching-off time, and its Wan and LTX-2 points come from an earlier denoiser program, as does LTX-2's quality; LTX-2's online-delta mean omits one prompt that skipped no step. Wan's device throttles during its 12-minute caching-off loop, which lifts its speedups past the ceiling; its quality is scored without calibrated adaptive.}
\label{fig:teacache}
\end{figure}

We evaluate the three caching modes of Section~\ref{sec:caching}: \emph{fixed cadence} at cadence~2; \emph{online-delta} at its default $\alpha = 0.6$, untuned; and \emph{calibrated adaptive}, with its curve fitted per model on prompts other than the evaluation prompts and its threshold set to cadence~2's skip budget. All modes keep 5 initial and 5 final full steps, so within a model they differ mainly in \emph{which} steps they cache. As a further baseline, \emph{fewer steps} runs without caching at the number of denoiser calls cadence~2 executes. Figure~\ref{fig:teacache} reports the speedups and the quality.

\paragraph{Findings.}
\emph{Caching saves the skipped denoiser calls, not time per call.} An executed call costs the same in every mode---830--845\,ms on HunyuanVideo, 835\,ms on Qwen-Image---and a cached step costs 0.3--7\,ms. The speedup is therefore set by the fraction of calls skipped. Because the initial and final full steps are fixed, official step counts skip proportionally more than short runs: cadence~2 skips 20 of 50 steps for a ceiling of 1.67$\times$ and reaches 1.66--1.68$\times$; on LTX-2 it skips 15 of 40 for a ceiling of 1.60$\times$. Wan is the exception, for a reason outside the cache: its caching-off loop runs for 12 minutes, during which the device throttles each executed call from 6.4 to 7.4\,s, so a run with fewer calls is also faster per call and cadence~2 reaches 1.76$\times$. Running fewer steps with the same number of calls takes the same time to within 0.5\% on every model, Wan included, so between caching and fewer steps the question is quality alone.

\emph{At the same number of denoiser calls, caching keeps far more of the output than sampling fewer steps.} On every model, cadence~2 is 7.4--11.6\,dB above fewer steps in mean PSNR against the uncached output (Figure~\ref{fig:teacache}), and every caching mode is at least 5.0\,dB above it; on the four models scored by LPIPS, the perceptual distance agrees, with mean LPIPS 0.05--0.15 under caching against 0.19--0.27 under fewer steps. The advantage holds prompt by prompt: caching beats fewer steps on 183 of those four models' 192 prompt--mode pairs and on all 10 of Wan's. Caching keeps the official noise schedule and reuses an output only where the trajectory changes slowly, whereas fewer steps coarsens the schedule everywhere. The worst prompt still loses much more than the mean, down to 15.5--17.8\,dB on the image models and HunyuanVideo and 17.0\,dB on Wan, so the mean alone overstates what every prompt keeps.

\emph{No mode is best everywhere, and online-delta's skip budget depends on the model and the prompt.} Cadence~2 keeps the most on FLUX, Qwen-Image, and, by 0.7\,dB, Wan, online-delta on HunyuanVideo and LTX-2, and calibrated adaptive is never the best, losing up to 3.0\,dB to cadence~2 at the same skip budget on HunyuanVideo. At the default $\alpha$, online-delta skips 18--20 of 50 steps on the 50-step models' timing prompts but only 8 of 40 on LTX-2's (1.25$\times$ against cadence~2's 1.60$\times$). Across the 16 prompts its budget varies by model: 19--20 skips on FLUX, 18--20 on HunyuanVideo, 13--19 on Qwen-Image, and 0--15 on LTX-2, where one prompt skips none at all; on Wan's five prompts it skips 17--19. The mode spends its budget where the output changes slowly, which buys LTX-2 its highest quality (37.3\,dB) at a speed that is not known in advance.

\begin{table}[t]
\caption{Signal of calibrated adaptive caching: the probe's cost per scheduler step and how well the fitted curve predicts the output change ($R^2$ over 8 recording prompts).}
\label{tab:cache-boundary}
\centering
\footnotesize
\setlength{\tabcolsep}{4pt}
\begin{tabular}{@{}llrr@{}}
\toprule
model & device probe & ms/step & fit $R^2$ \\
\midrule
FLUX.1-dev   & fused in the denoiser & 2.3  & 0.60 \\
Qwen-Image   & fused in the denoiser & 7.3  & 0.68 \\
Wan~2.1      & separate program  & --$^{o}$ & 0.13 \\
HunyuanVideo & separate program  & 17.8 & 0.71 \\
LTX-2        & separate program  & 2.6  & 0.82 \\
\bottomrule
\end{tabular}
\par\smallskip
{\raggedright\scriptsize $^{o}$Not resolvable: Wan's adaptive run executed its steps faster than its caching-off run, which swamps a per-step cost.\par}
\vskip -0.1in
\end{table}

\emph{Calibrated adaptive is decided by how well its signal predicts, not by what the signal costs.} Every model runs the probe on the device (Table~\ref{tab:cache-boundary}), including HunyuanVideo, whose probe loads beside its denoiser at this shape. The probe costs 2--18\,ms per step, 0.6--2\% of a denoiser call, so at the same skip count the mode lands within 4\% of cadence~2's time. What separates the models is the fit: the curve explains 60--82\% of the variance in output change on four models but 13\% on Wan, where a timing gain is no evidence that the mode skips the right steps.

\subsection{Serving}
\label{sec:serving-eval}

Section~\ref{sec:serving} claims that a restart costs a load and not a compile, and that recovery after a timeout or crash leaves no state behind. We test both on the model process of \texttt{difflet serve}, and measure request latency under open-loop load.

\begin{table}[t]
\caption{Startup of one resident model at TP=4, until the server reports ready. Recompiles counts the programs rebuilt when switching from caching off to calibrated adaptive.}
\label{tab:startup}
\centering
\footnotesize
\setlength{\tabcolsep}{4pt}
\begin{tabular}{@{}lrrrr@{}}
\toprule
& first start & \multicolumn{2}{c}{restart (s)} & \\
\cmidrule(lr){3-4}
model & (min) & cold & warm & recompiles \\
\midrule
FLUX.1-dev   & 18.5  & 340 & 91  & 1 of 1 \\
Qwen-Image   & 34.8  & 640 & 129 & 1 of 3 \\
Wan~2.1      & 66.9$^{v}$ & 1{,}102$^{u}$ & 786$^{u}$ & --$^{p}$ \\
HunyuanVideo & 116.8 & 624 & 194 & --$^{p}$ \\
LTX-2        & 33.8  & 811 & 155 & --$^{p}$ \\
\bottomrule
\end{tabular}
\par\smallskip
{\raggedright\scriptsize $^{p}$Resident video serving exposes no step-caching policy. $^{u}$Includes the startup smoke request, whose 81-frame host decode takes about 510\,s. $^{v}$Builds no VAE, since the model process decodes on the host; the standalone build with the chunked VAE takes 75.1\,min (Table~\ref{tab:e2e}).\par}
\vskip -0.1in
\end{table}

\begin{table}[t]
\caption{Request latency under open-loop Poisson load through \texttt{difflet serve} over HTTP: for FLUX, one TP4 server against two TP2 servers on separate core pairs, each request sent to the server with fewer requests in flight; for HunyuanVideo, one TP4 server (its TP2 denoiser does not fit in HBM). Both configurations receive the same arrival sequence; 20 requests per load level. Load $\rho$ is the realized arrival rate times the isolated TP4 request latency (FLUX 14.2\,s; HunyuanVideo 58.5\,s in this run, 59.9\,s in Table~\ref{tab:e2e}). SLO: completion within 2.1$\times$ that latency (30\,s and 123\,s).}
\label{tab:serving-load}
\centering
\footnotesize
\setlength{\tabcolsep}{3pt}
\begin{tabular}{@{}lrrrr@{}}
\toprule
& \multicolumn{2}{c}{p50 / p99 (s)} & \multicolumn{2}{c}{SLO met} \\
\cmidrule(lr){2-3}\cmidrule(lr){4-5}
FLUX, $\rho$ & TP4 & DP2$\times$TP2 & TP4 & DP2$\times$TP2 \\
\midrule
0.78 & 39.4 / 65.3  & 39.1 / 70.0  & 40\% & 30\% \\
0.99 & 33.0 / 53.8  & 46.7 / 69.3  & 40\% & 25\% \\
1.18 & 65.9 / 105.3 & 74.7 / 110.7 & 20\% & 20\% \\
\midrule
\rlap{HunyuanVideo, $\rho$} & TP4 & & TP4 & \\
\midrule
0.81 & 136.1 / 244.2 & n/a & 45\% & n/a \\
0.95 & 89.4 / 176.3  & n/a & 80\% & n/a \\
1.16 & 201.1 / 374.2 & n/a & 20\% & n/a \\
\bottomrule
\end{tabular}
\vskip -0.1in
\end{table}

\paragraph{Findings.}
\emph{A restart costs a load, not a compile---unless the probe changes.} The first start of a serving profile compiles its programs and takes 18--117\,min (Table~\ref{tab:startup}). Every later start validates them and pays the weight load, which the host page cache bounds: 91--786\,s warm, 340--1{,}102\,s cold. Changing the caching target recompiles nothing, but switching from caching off to calibrated adaptive rebuilds the program that carries the probe: FLUX's whole pipeline program (1{,}353\,s) and Qwen-Image's denoiser (1{,}419\,s).

\emph{Recovery leaves no state behind.} We test both failure paths on FLUX, repeating one prompt and seed; two reference requests produce byte-identical images. A 5\,s request timeout returns an error at 5.0\,s, and the same model process serves again 9.3\,s later, once its running denoising loop reaches the next cancellation point. Killing the model process with SIGKILL mid-request returns an error after 0.6\,s; \name spawns a new one, reloads the weights, and is ready 65.6\,s after the kill. After both faults the same seed reproduces the reference image byte for byte, so no sampling or caching state survives a request.

\emph{Under load, splitting the chip into replicas does not pay for resident FLUX.} Table~\ref{tab:serving-load} offers both configurations the same Poisson arrivals. At no load level does DP2$\times$TP2 beat TP4: their median latency ties at $\rho = 0.78$, and at $\rho = 0.99$ TP4's median is 33.0\,s against 46.7\,s, with the 30\,s SLO met for 40\% of requests against 25\%. Each TP2 replica needs 26--27\,s for a request that TP4 serves in 14.2\,s, so a request that finds a replica idle already spends most of the SLO, while the second replica adds only the 5\% of throughput measured in Section~\ref{subsec:parallelism-eval}. Both configurations complete the same number of requests per hour at every level, because arrivals, not capacity, limit them. Queueing dominates both: past $\rho \approx 1$ the 99th percentile exceeds 100\,s for either configuration. With 20 requests per level these percentiles are coarse---TP4's median is higher at $\rho = 0.78$ than at $\rho = 0.99$ (39.4 against 33.0\,s)---but at every level TP4 has the lower 99th percentile and meets the SLO at least as often. HunyuanVideo, whose 58.5\,s request leaves no room for a second replica, queues the same way: past $\rho \approx 1$ its 99th percentile reaches 374\,s, 6.4$\times$ the isolated latency, and only 20\% of requests meet the SLO. Its $\rho = 0.81$ level queues more than its $\rho = 0.95$ level (median 136 against 89\,s), an artifact of drawing only 20 arrivals per level.

\section{Related Work}
\label{sec:related}

\paragraph{Reducing denoising computation.} Fast solvers~\cite{dpmsolver}, progressive distillation~\cite{progdistill}, and consistency models~\cite{consistency} reduce the number of denoiser evaluations needed for generation. Caching instead reuses computation across timesteps: DeepCache and $\Delta$-DiT reuse intermediate features~\cite{deepcache,deltadit}, while PAB and FasterCache exploit redundancy in attention and guidance computations~\cite{pab,fastercache}. TeaCache uses timestep-modulated input changes to guide reuse~\cite{teacache}. TaylorSeer predicts future features from their history~\cite{taylorseer}, and CG-Taylor uses early-block prediction error to decide when to recompute~\cite{cgtaylor}. \name builds on temporal reuse; what it adds is how to obtain the reuse signal, and decide each step, when the denoiser is a program compiled ahead of time.

\paragraph{Parallel inference systems.} Ulysses redistributes sequence shards into attention-head shards~\cite{ulysses}, while Ring Attention circulates key-value blocks among devices~\cite{ringattention}. DistriFusion reuses previous-step features to overlap communication with computation~\cite{distrifusion}; PipeFusion combines patch and layer partitioning into a pipeline~\cite{pipefusion}. xDiT composes parallel modes~\cite{xdit}, while FastVideo integrates sparse attention and distillation~\cite{fastvideo}. On specialized accelerators, MaxDiffusion provides JAX implementations for TPUs~\cite{maxdiffusion}, and AoiZora plans sharding and placement against TPU topology~\cite{aoizora}. NxD Inference supports FLUX on Trainium~\cite{nxdi}. \name instead runs five image and video pipelines on Trainium through one backend, and splits their joint attention in a layout that Trainium's fused kernels accept.

\paragraph{Diffusion serving.} DiffServe routes requests through cascades of diffusion models and adjusts resource allocation to query demand~\cite{diffserve}. NIRVANA reuses intermediate states across similar requests~\cite{nirvana}, while TridentServe jointly plans stage placement and request dispatch~\cite{tridentserve}. These systems optimize how requests share model and accelerator resources. \name works below these systems, inside one Trainium model process: it keeps programs loaded and validated across requests and recovers from a timeout without letting two generations overlap.

\section{Conclusions}

On Trainium, a diffusion model runs as fixed-shape programs compiled ahead of time, while its pipeline, prompts, and caching decisions keep changing. \name runs each model's pipeline on the host over one compilation backend, splits joint attention across cores so that it matches unsplit attention, and skips denoising steps under host control. Device probes keep comparison state on Trainium and return one score per step to the host, while the resident model process keeps every program loaded across requests. Together, these mechanisms provide a common inference and serving framework for five image and video model families. The broader lesson is to decide early what each program fixes and what the host may still change: \name compiles shapes, parallel configurations, and kernels once, and exchanges per-request and per-step information with the device as small runtime values---a count of valid text keys going in, one change score per step coming out.

\bibliography{ref}

\bibliographystyle{mlsys2025}

\appendix

\end{document}